\documentclass[conference]{IEEEtran}
\IEEEoverridecommandlockouts

\usepackage{cite}
\usepackage{amsmath,amssymb,amsfonts}
\usepackage{algorithmic}
\usepackage{graphicx}
\usepackage{textcomp}
\usepackage{xcolor}
\def\BibTeX{{\rm B\kern-.05em{\sc i\kern-.025em b}\kern-.08em
    T\kern-.1667em\lower.7ex\hbox{E}\kern-.125emX}}

\newcommand{\ModelName}{3T2-SASV} 
\usepackage{url}
\usepackage{comment}
\usepackage{multirow}
\usepackage{adjustbox}
\usepackage{balance}
\usepackage{booktabs, subcaption}
\usepackage{hyperref}

\begin{document}

\title{Integrated Spoofing-Robust Automatic Speaker Verification via a Three-Class Formulation and LLR}


\author{
\IEEEauthorblockN{
Kai Tan\textsuperscript{1,*},
Lin Zhang\textsuperscript{1,*},
Ruiteng Zhang\textsuperscript{2},
Johan Rohdin\textsuperscript{3},\\
Leibny Paola Garc\'ia\textsuperscript{1},
Zexin Cai\textsuperscript{1},
Sanjeev Khudanpur\textsuperscript{1},
Matthew Wiesner\textsuperscript{1},
Nicholas Andrews\textsuperscript{1}
}

\IEEEauthorblockA{
\textsuperscript{1}\textit{HLTCOE, CLSP, Johns Hopkins University},
Baltimore, MD, USA\\
\textsuperscript{2}\textit{Tianjin University},
Tianjin, China  
\textsuperscript{3}\textit{Speech@FIT, Brno University of Technology},
Brno, Czechia\\
\texttt{ktan17@jh.edu, zlin@ieee.org}\\
\textsuperscript{*}Equal contribution.
}
}

\maketitle

\begin{abstract}

Spoofing-robust automatic speaker verification (SASV) is challenging because the system must accept bona fide target trials while rejecting both bona fide nontarget and spoofed trials. Existing integrated SASV methods often collapse rejection evidence into a single score, making rejection causes less interpretable and prior adjustment difficult. We propose \ModelName, an end-to-end SASV framework based on a cross-encoder and a three-class formulation. Given an enrollment-test pair, the model predicts target, nontarget, and spoof logits, which are transformed into an SASV log-likelihood ratio (LLR) for the final accept/reject decision. By explicitly preserving class-level evidence, the proposed framework enables post-hoc operating-prior adaptation and clearer rejection analysis. Experiments demonstrate competitive performance on ASVspoof~5 and improvements on SpoofCeleb over existing integrated baselines. Visualization and score analysis further show a clearer separation of rejection patterns. Additional results show that the framework generalizes well to SSL-based embedding extractors.
\end{abstract}


\section{Introduction}
\label{sec:introduction}
{Automatic speaker verification (ASV) aims to verify whether a test utterance matches a claimed speaker identity. 
Although ASV provides a convenient authentication mechanism for security-sensitive applications, it remains vulnerable to spoofing attacks~\cite{WU2015130}, particularly those generated by advanced voice conversion (VC) and text-to-speech (TTS) systems~\cite{peng-etal-2024-voicecraft, Zhou2025IndexTTS2AB}.
Speaker mismatch and spoofing are conventionally addressed by ASV and countermeasure (CM)~\cite{evans2013spoofing}, respectively. In a practically reliable system, only bona fide target trials should be accepted, while both bona fide nontarget trials and spoofed trials should be rejected. This motivates spoofing-robust automatic speaker verification (SASV), which integrates speaker verification and spoof detection into a unified trial-level decision~\cite{jung22c_interspeech}. 
To encourage research in this area, several SASV challenges and benchmarks have been organized~\cite{jung22c_interspeech,ASVspoof5,wu2025wildspoof}.

Existing SASV approaches can be categorized into two groups. The first one is fusion-based approaches, as illustrated in Figure~\ref{fig:propose_vs_exist} (a). ASV and CM are trained independently, and fused at either the score level~\cite{Zhang2022, zeng22_interspeech, zeng2024joint, wang24linterspeech} or the embedding level~\cite{Buker2025ParameterSharing}. These systems typically retain separate ASV and CM evidence before fusion, resulting in a double-score or two-pathway decision interface. The fusion module can range from non-trainable rules, such as averaging or weighted summation to probabilistic back-ends, such as Gaussian back-end fusion~\cite{todisco2018integrated}, and discriminatively trained score-fusion models, including logistic-regression-based LLR optimization~\cite{rohdin24asvspoof} and a-DCF-oriented back-end optimization~\cite{Kurnaz2025JointOO}.
Although effective, these modular approaches remain dependent on the compatibility between their components and may require joint adaptation or re-tuning under emerging spoofing attacks or changing operating conditions~\cite{mun23_interspeech}.


\begin{figure*}
    \centering
    \includegraphics[width=0.98\textwidth]{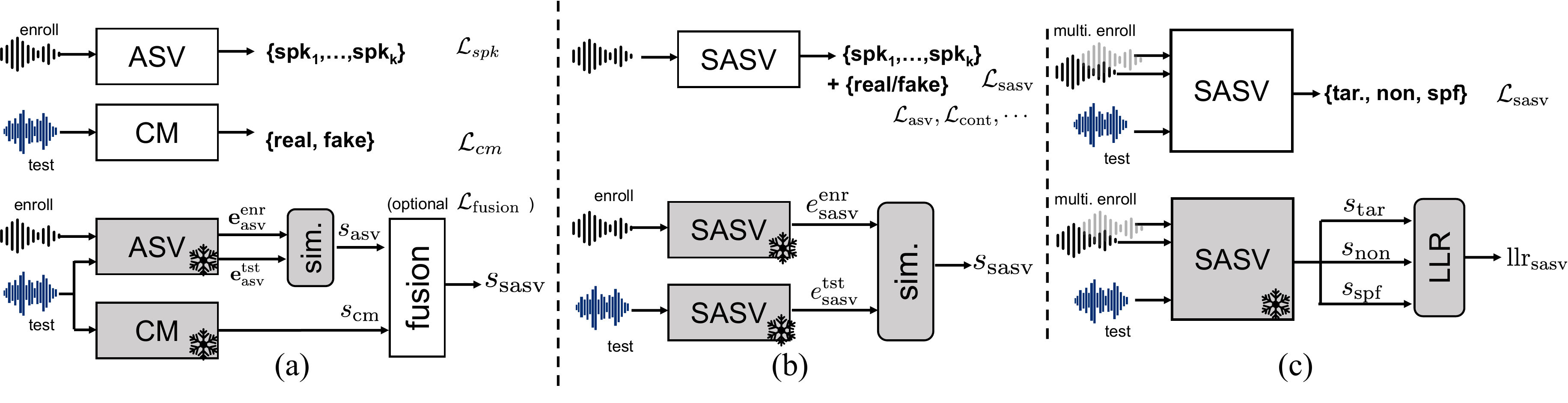}
    \vspace{-3mm}
    \caption{
    Comparison of SASV paradigms. 
    (a) Fusion-based SASV typically trains separate ASV and CM using $\mathcal{L}_{\mathrm{spk}}$ and $\mathcal{L}_{\mathrm{cm}}$, and may include a fusion loss $\mathcal{L}_{\mathrm{fusion}}$ in either score level (shown here) or embedding level. 
    (b) Existing bi-encoder-based integrated SASV systems use a unified SASV objective $\mathcal{L}_{\mathrm{sasv}}$, often together with ASV-related or contrastive objectives, and independently encode enrollment and test utterances before scoring. 
    (c) The proposed \ModelName~adopts a cross-encoder architecture with a three-class formulation and converts the class logits into an SASV LLR for inference. 
    Unlike the double- or single-score formulations in (a) and (b), the proposed method preserves explicit class-level evidence for the three SASV hypotheses. Gray boxes denote non-trainable modules.
    }
    \vspace{-6mm}
    \label{fig:propose_vs_exist}
\end{figure*}

{The second group is integrated SASV systems, as illustrated in Figure~\ref{fig:propose_vs_exist}(b). These methods directly learn a unified SASV representation or decision function for final SASV scoring. Although they may employ multi-task or auxiliary objectives during training, their components are jointly optimized toward a single integrated SASV objective. In this paradigm, enrollment and test utterances are processed within the same integrated SASV framework.
In 2022, Teng \MakeLowercase{\textit{et al.}}~\cite{teng22_interspeech} introduced SA-SASV, an end-to-end spoof-aggregated framework that combines a pretrained ASV encoder with an auxiliary raw-waveform encoder. It employs multi-task classifiers and spoof-source-aware objectives to organize bona fide and spoofed utterances in a shared SASV representation space. In the same year, Kang \MakeLowercase{\textit{et al.}}~\cite{kang22_interspeech} proposed an end-to-end framework with a spoof-aware contrastive objective. This objective encourages embeddings from bona fide target utterances to be close while separating them from nontarget or spoofed utterances.
Subsequently, Mun~\MakeLowercase{\textit{et al.}}\cite{mun23_interspeech} introduced an improved integrated SASV model. In addition to the spoof-aware contrastive objective, their method incorporates detailed speaker identification and ``real or fake'' labels into training. }

Although the latter method~\cite{mun23_interspeech} improves performance over~\cite{kang22_interspeech} to some extent, it introduces additional training objectives that may not be necessary for SASV. Moreover, existing integrated methods adopt a bi-encoder design~\cite{Humeau2019PolyencodersAA}, where enrollment and test utterances are encoded separately and compared only through a similarity score. This limits explicit interaction between the two inputs and makes it difficult to tell whether a rejection is caused by speaker mismatch or spoofing. More importantly, a single SASV score obtained through score fusion does not naturally support separate prior adjustment for the three underlying trial classes. 

A simpler yet effective framework supporting class-specific prior adjustment is therefore needed. We argue that a desirable SASV criterion should preserve three-class evidence so that different operating priors can be applied at inference. Existing integrated systems mentioned above do not fully align their objectives with the three SASV trial classes: target, nontarget, and spoofed trials. As a result, their outputs do not explicitly support separate prior adjustment or a principled SASV LLR. In contrast, a three-class formulation preserves two rejection classes and better matches the SASV decision space.

{Therefore, we propose \ModelName, a cross-encoder-based integrated SASV framework that combines a three-class formulation with LLR-based inference, as shown in Fig.~\ref{fig:propose_vs_exist} (c). 
The proposed framework (1) takes enrollment--test pairs as input and explicitly models their interaction through a cross-attention architecture, (2) employs a trial-level objective to distinguish target, nontarget, and spoofed trials, and (3) converts the \textbf{3}-class logits \textbf{T}o a \textbf{2}-decision LLR between the accept and reject hypotheses. 
By aligning the model outputs with the three SASV classes, the proposed formulation preserves separate speaker-mismatch and spoofing-related evidence, enables post-hoc adjustment of class priors, and provides a principled derivation of the SASV LLR. Despite its simple design, the proposed framework achieves competitive performance on ASVspoof~5 and superior performance on SpoofCeleb compared with existing integrated SASV systems. 

\section{Proposed method}
In this section, we introduce \ModelName, a cross-encoder-based integrated SASV framework. Given an enrollment–test pair as input, the proposed framework produces three-class logits corresponding to target, nontarget, and spoof class, and converts them into an SASV LLR for the final binary decision.

\subsection{Task definition} 
\label{sec:task_reformulation}
SASV~\cite{jung22c_interspeech} typically assumes that all enrollment utterances are bona fide and that each enrollment-test trial belongs to one of three classes: (1) Target: the test utterance is bona fide and spoken by the enrolled target speaker, (2) Nontarget: the test utterance is bona fide but spoken by a nontarget speaker, and (3) Spoof: the test utterance is spoofed.
Given these three cases, SASV makes a binary decision between:
\begin{itemize}
    \item $\mathcal{H}_A$: Accept hypo. when the trial is target.
    \item $\mathcal{H}_R$: Reject hypo. when the trial is nontarget or spoof.
\end{itemize}

\begin{figure*}
    \centering
    \includegraphics[width=0.95\textwidth, clip]{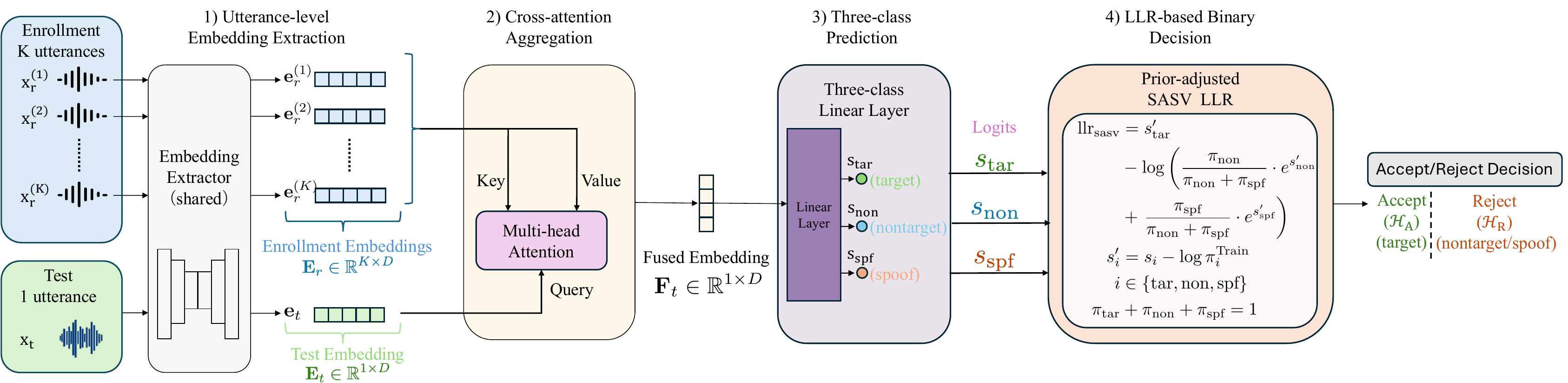}
    \caption{Overview of the proposed \ModelName~framework. Given enrollment and test utterances, the embedding extractor first produces utterance-level representations, which are then aggregated to obtain a fused representation $\mathbf{F}_t$. The three-class classifier predicts target, nontarget, and spoof logits, which are converted into an SASV log-likelihood ratio $\mathrm{llr}_{\mathrm{SASV}}$ for the final accept/reject decision.}
    \label{fig:propose_method}
    \vspace{-5mm}
\end{figure*}

\subsection{Cross-encoder SASV with three-class objective}\label{sec:3-class_model}
As introduced in \S\ref{sec:introduction}, most existing integrated SASV frameworks produce a single binary score for decision, making it difficult to trace the reason for rejection, and not allowing prior adjustment. Based on the \S\ref{sec:task_reformulation}, it is natural to \textit{model SASV as a three-class classification} problem over target, nontarget, and spoofed trials. However, this formulation received limited attention in existing integrated SASV studies.
The proposed \ModelName~framework is illustrated in Fig.~\ref{fig:propose_method}. It takes an enrollment-test pair as input and consist of four stages: utterance-level embedding extraction, cross-attention aggregation, three-class prediction, and LLR-based binary decision. The following subsections describe each stage in detail.

\subsubsection{Hard-pair enrollment-test construction}\label{sec:hard-pair}

As a cross-encoder-based SASV framework, \ModelName~accepts enrollment-test trials for training. Motivated by the effectiveness of hard-pair mining in ASV~\cite{ren2025canllm}, we construct hard-pair trials using the available metadata.

For target trials, the enrollment and test utterances are selected from the same speaker. Whenever possible, the two utterances contain different linguistic content, thereby increasing the diversity of non-speaker attributes within target trials. For nontarget trials, the enrollment and test utterances are selected from different speakers while sharing as many non-speaker attributes as possible (e.g.: gender), forcing the model to learn speaker-related representations. For spoofed trials,
we select spoof utterances that match the enrolled speaker identity.

\subsubsection{Utterance embedding extraction}
Given $K$ enrollment utterances $\{\mathbf{x}_r^{(k)}\}_{k=1}^{K}$ and one test utterance $\mathbf{x}_t$, \ModelName~extracts enrollment embedding $\mathbf{E}_r \in\mathbb{R}^{K\times D}$ and test embedding $\mathbf{E}_t \in \mathbb{R}^{1\times D}$.

\subsubsection{Cross-attention aggregation}
\label{sec:fusion strategy}
We utilize multi-head cross-attention to aggregate $\mathbf{E}_r$ and $\mathbf{E}_t$ to a trial-level fused representation $\mathbf{F}_t\in\mathbb{R}^{1\times D}$, which was shown effective for aggregating multiple enrollment utterances in ASV~\cite{zeng2022attention}. Specifically, $\mathbf{E}_t$ is used as the query, while $\mathbf{E}_r$ is used as the key and value.
This design yields a test-conditioned representation that attentively aggregates multiple enrollment utterances from the same target speaker.

\subsubsection{Three-class objective}\label{sec:softmax_loss}
The proposed \ModelName~employs a three-class objective at the trial level. 
%
Given $\mathbf{F}_t$, a linear layer maps $\mathbf{F}_t$ to three logits $\mathbf{s}=[s_{\mathrm{tar}},s_{\mathrm{non}},s_{\mathrm{spf}}]$, corresponding to target, nontarget, and spoof class, respectively. The logits are then converted into posterior probabilities using the softmax function, and the model is trained with a three-class cross-entropy objective. 
\subsection{From three-class logits to binary decision (3T2) via LLR}
\label{sec:llr}
We convert the three-class prediction from \ModelName~to a binary accept-reject decision for SASV. 
An LLR-based framework~\cite{rohdin24asvspoof} was shown to be efficient to derive a general SASV LLR from Bayes decision theory. 
We follow this LLR-based idea, but derive the SASV LLR directly from the three-class logits of a single integrated model. 

This design preserves class-level evidence for the final binary decision, \textit{making it possible to distinguish whether a rejection is mainly due to speaker mismatch or spoofing. It also enables post-hoc adjustment of class priors}, allowing the decision rule to be adapted to different evaluation or application conditions without retraining the model.


\subsubsection{SASV LLR definition}
Given the enroll-test pair $\mathbf{x} = (\mathbf{x}_r, \mathbf{x}_t)$, the LLR in this study is defined as:
\begin{equation}
\begin{aligned}
    \text{llr}_\text{sasv}
    &= \log\frac{P(\mathbf{x} \mid \mathcal{H}_A)}{P(\mathbf{x} \mid \mathcal{H}_R)} \\
                  &= \log\frac{P(\mathbf{x} \mid \text{target})}{P(\mathbf{x} \mid \text{non}) \cdot \frac{\pi_{\text{non}}}{\pi_{\text{non}} + \pi_{\text{spf}}} + P(\mathbf{x} \mid \text{spf}) \cdot \frac{\pi_{\text{spf}}}{\pi_{\text{non}} + \pi_{\text{spf}}} },
\end{aligned}
\label{eq:llr_def}
\end{equation}
where $\mathcal{H}_A$ and $\mathcal{H}_R$ denote the accept hypothesis and reject hypothesis, respectively, as defined in \S\ref{sec:task_reformulation}.
The terms \( \pi_{\text{non}} \) and \( \pi_{\text{spf}} \) represent the prior probabilities of the nontarget and spoof classes.

\subsubsection{Converting three-class predictions to a binary decision}
Given the posterior probabilities $P(y=i \mid \mathbf{x})$ obtained by applying the softmax function to the model logits as mentioned in \S\ref{sec:softmax_loss}, the corresponding class-conditional likelihood can be expressed via Bayes' theorem as:
\begin{equation}
\begin{aligned}
    P(\mathbf{x} \mid y=i) &= \frac{P(y=i \mid \mathbf{x}) \cdot P(\mathbf{x})}{P(y=i)}, \\
    \log P(\mathbf{x} \mid y=i) &= s_i - \log \pi^\text{Train}_i + \text{const}.
\end{aligned}
\end{equation}
Here, $\pi^\text{Train}_i = P(y=i)$ is the prior probability of the class $i$ in the training set, and $\text{const}$ refers to the constant term ($P(\mathbf{x})$) across all classes, which does not affect LLR computation.
%
We then adjust the logits to remove effects by the prior $\pi^\text{Train}$ in the training set:
%
\begin{equation}
    s'_i = s_i - \log \pi^\text{Train}_i,
\label{eq:adjusted_logits}
\end{equation}
Here, $\pi^\text{Train}_i  = \frac{N_i}{N}$. 
Then the LLR can be computed using the adjusted logits:
\begin{equation}
\text{llr}_\text{sasv} = s'_\text{tar} - \log\left( \frac{\pi_{\text{non}}}{\pi_{\text{non}} + \pi_{\text{spf}}} \cdot e^{s'_\text{non}} + \frac{\pi_{\text{spf}}}{\pi_{\text{non}} + \pi_{\text{spf}}} \cdot e^{s'_\text{spf}} \right),
\label{eq:weighted_llr}
\end{equation}
Note that $\pi_{\text{tar}} + \pi_{\text{non}} + \pi_{\text{spf}} = 1$, and the rejection hypothesis prior is defined as $\pi_{\text{non}} + \pi_{\text{spf}}$.

It should be noted that to accurately compute the SASV LLR in Eq.~\ref{eq:adjusted_logits}, the logits, $s'$, need to be calibrated. The fact that the min a-DCF ignores calibration of the SASV LLR does not remove the need for this. In principle, since the three-class classifier is trained by cross-entropy loss, the obtained logits should be well calibrated. However, in practice, there are reasons to attempt an additional calibration step. First, scores are rarely well calibrated across their whole range, and the multi-class training may emphasize calibration in a different range than what is important for the SASV LLR. Second, if there is a domain shift between the training data for the three-class classifier, calibration with a small amount of in-domain development data can be very beneficial. In our initial experiments, we therefore explored an additional calibration step using affine transformations trained with binary cross-entropy for the SASV decision. However, we did not see an obvious benefit of this when evaluated using min a-DCF. So we do not include this step in the remaining experiments.

\section{Experimental Setup}
\subsection{Dataset and training trials}
We conduct experiments mainly on two datasets:
\begin{itemize}
    \item ASVspoof~5~\cite{ASVspoof5} is built upon crowdsourced speech from Multilingual LibriSpeech~\cite{pratap20_interspeech} with speaker-disjoint training/development/evaluation partitions and provides bona fide and spoofed utterances with multiple attack types. We use ASVspoof~5 Track 2 for this study.
    \item SpoofCeleb~\cite{10839331} is a large-scale in-the-wild dataset designed for both SDD and SASV. It is derived from VoxCeleb1~\cite{Nagrani2020Voxceleb:Wild} via an automated pipeline that produces a TTS-trainable subset (TITW-Easy~\cite{jung2025texttospeechsynthesiswild}) and then synthesizes spoofed speech using 23 TTS systems. 
\end{itemize}

Since the above-mentioned datasets do not provide predefined training trials, we construct 12,000 hard-pair training trials for the proposed method as described in \S\ref{sec:hard-pair}. 
We randomly determine the number of times each speaker serves as the reference speaker, subject to the constraint that every speaker appears at least once. For each reference speaker, trials are formed using bona fide, nontarget, and spoof samples with a balanced class ratio of 1:1:1.

We apply hard-pair sampling to ASVspoof~5, as it yields better performance on development set(min a-DCF of 0.21 with hard-pair sampling versus a-DCF of 0.25 with random sampling). For SpoofCeleb, however, we use random sampling rather than hard-pair sampling as it provides limited metadata.

\subsection{Metrics}
System performance is assessed using the architecture-agnostic Detection Cost Function (a-DCF)~\cite{shim2024adcf}, following the ASVspoof~5 evaluation protocol. Lower a-DCF indicates better performance.
Unless otherwise specified, we use the class priors adopted in ASVspoof~5 and SpoofCeleb: $\pi_{\text{tar}} = 0.9405,\quad \pi_{\text{non}} = 0.0095,\quad \pi_{\text{spf}} = 0.0500.$

\subsection{Experimental Configuration}\label{sec:config}

\subsubsection{Baselines with conventional backbones}
Following the open-source integrated SASV baseline~\cite{mun23_interspeech}, we adopt two conventional backbone networks as utterance embedding extractors: MFA-Conformer~\cite{zhang22h_interspeech} and SKA-TDNN~\cite{SKA-TDNN}. For each backbone, we compare \ModelName~with two integrated SASV baselines: Integrated B04~\cite{ASVspoof5,mun23_interspeech} and SA-SASV~\cite{teng22_interspeech}. Unless otherwise stated, experiments on ASVspoof~5 follow the ASVspoof~5 B04 baseline configuration~\cite{ASVspoof5} (ASV5-B04), while experiments on SpoofCeleb follow the configuration reported in the original SpoofCeleb paper~\cite{10839331}.

Since the original SA-SASV system was developed using a different encoder, dataset, and evaluation metric, directly comparing its originally reported results with those of \ModelName~would not provide a controlled comparison. We therefore reimplement SA-SASV under the same experimental conditions as the proposed method while retaining its principal SASV training formulation. Specifically, the reimplemented systems use the same pretrained ASV initialization, MFA-Conformer and SKA-TDNN as the utterance embedding extractor and share the same training partitions, input duration, model initialization, and min a-DCF evaluation protocol as the other systems. This controlled setting allows the comparison to focus on differences in the SASV formulation rather than differences in the backbone or evaluation protocol.

\subsubsection{Proposed method with conventional extractor}
The proposed \ModelName~is implemented using the same MFA-Conformer and SKA-TDNN backbones to ensure comparability with the corresponding baselines. The implementation is released on GitHub\footnote{\url{https://github.com/Ka1Kevin/3T2-SASV}}.

For fair comparison, \ModelName~with conventional extractors is initialized from the pretrained ASV checkpoint as the baselines. Specifically, we use the Stage-I pretrained model\footnote{\url{https://github.com/sasv-challenge/SASV2_Baseline}} released by Mun \MakeLowercase{\textit{et al.}}~\cite{mun23_interspeech}, which is also used in ASV5-B04.

\subsubsection{Proposed method with SSL extractor}
In addition to conventional acoustic backbones, we further evaluate Self-Supervised Learning (SSL)-based frontends to examine whether the proposed framework generalizes across different speech representation models. Specifically, we use pretrained wav2vec~2.0 Base and wav2vec~2.0 Large~\cite{wav2vec2}, following the SSL settings supported in WeDefense~\cite{wedefense}\footnote{\url{https://huggingface.co/s3prl/converted_ckpts/resolve/main/}}. 
Unlike MFA-Conformer and SKA-TDNN, which operate on hand-crafted acoustic features, wav2vec~2.0 directly processes raw waveforms and produces contextualized speech representations. These representations are then fed into the subsequent aggregation and three-class classification modules. All components, including the pretrained wav2vec~2.0 encoder,
utterance-level backend, cross-attention module, and three-class classifier,
are jointly fine-tuned using the SASV training objective.

\subsubsection{Other details}
Based on preliminary experiments on the development set, all experiments use one enrollment utterance per trial (K=1)\footnote{We chose K=1 based on ablation study on the dev. set. Although K=1 does not perform query-dependent selection, we use $F_t = \text{LN}(E_t + \text{Att}(E_t, E_r, E_r))$, so the trial rep. includes both enroll. and test information.}, crop or zero-pad all utterances to 2 seconds, and the batch size is set to 16. No data augmentation is applied. For the cross-attention module described in \S~\ref{sec:fusion strategy}, we use a single multi-head attention layer with an embedding dimension of 192, 4 attention heads, and a dropout rate of 0.1. The attention output is combined with the test embedding through a residual connection, followed by Layer Normalization. No feed-forward layer or additional stacked attention block is used. More results on the selection of key parameters ($K$, utterance length, number of attention heads, hard pairs, and aggregation), are provided in the appendix\footnote{\url{https://arxiv.org/abs/2603.13780}}.

Following~\cite{mun23_interspeech}, for both the baselines and the proposed method using conventional backbones, the cropped waveforms are converted into 80-dimensional log Mel-filterbank features using a 25~ms window and a 10~ms frame shift. All components are jointly updated during training.
For experiments with SSL extractors, the cropped raw waveforms are directly fed into the SSL frontends, and the complete framework is
fine-tuned end to end. 

Unless otherwise stated, the same optimization settings are used for all systems. The models are optimized using AdamW with a weight decay of $1\times10^{-4}$. The learning rate is linearly warmed up from zero to $1\times10^{-3}$ during the first 10 epochs and is then exponentially decreased to $5\times10^{-5}$. Training is conducted for a maximum of 50 epochs. Early stopping with a patience of 5 epochs is applied according to the development set min a-DCF.


\section{Results and Discussion}
\label{sec:results}

\subsection{Comparison with existing methods}
We compare the proposed \ModelName~with two integrated SASV baselines, ASV5-B04~\cite{mun23_interspeech} and SA-SASV~\cite{teng22_interspeech}, on ASVspoof~5 and SpoofCeleb. We reimplemented baselines (with comparable results 0.5882 vs. 0.5741 in ~\cite{ASVspoof5} w/o augmentation, 0.2914 vs. 0.2901 in ~\cite{10839331}), and ran each experiment three times with different random seeds, reporting the average results.
Results are summarized in Table~\ref{tab:comparison_with_baseline}. 

\begin{table}[t!]
\centering
\caption{Comparison with existing methods (min a-DCF on evaluation sets).}
\label{tab:comparison_with_baseline}
\footnotesize
\vspace{-0.15cm}
\begin{tabular}{llll}
\toprule
                  & Emb. Extractor & {Method}     & min a-DCF  \\
\midrule
\multirow{6}{*}{\rotatebox[origin=c]{90}{ASVspoof~5}}  & MFA-Conformer & ASV5-B04~\cite{ASVspoof5} & \multicolumn{1}{r}{0.5882} \\
                            & MFA-Conformer & SA-SASV~\cite{teng22_interspeech} & \multicolumn{1}{r}{0.6025} \\
                            & MFA-Conformer & Proposed \ModelName & \multicolumn{1}{r}{0.6029} \\
                            & SKA-TDNN      & ASV5-B04 & \multicolumn{1}{r}{0.5788} \\
                            & SKA-TDNN      & SA-SASV & \multicolumn{1}{r}{0.5830} \\
                            & SKA-TDNN      & Proposed \ModelName & \multicolumn{1}{r}{0.5961} \\
\midrule
\multirow{6}{*}{\rotatebox[origin=c]{90}{SpoofCeleb}} & MFA-Conformer & ASV5-B04 & \multicolumn{1}{r}{0.2384} \\
                            & MFA-Conformer & SA-SASV & \multicolumn{1}{r}{0.2643} \\
                            & MFA-Conformer & Proposed \ModelName & \multicolumn{1}{r}{0.1295} \\
                            & SKA-TDNN      & ASV5-B04 & \multicolumn{1}{r}{0.2914} \\
                            & SKA-TDNN      & SA-SASV & \multicolumn{1}{r}{0.3290}   \\
                            & SKA-TDNN      & Proposed \ModelName & \multicolumn{1}{r}{0.1423}   \\
                            
\bottomrule
\end{tabular}
\vspace{-0.3cm}
\end{table}


On ASVspoof~5, our method achieves performance comparable to the baselines across both embedding extractors. One possible explanation is that ASVspoof~5 includes strong spoofing attacks and challenging adversarial conditions, which limit the margin for improvement over existing integrated baselines. A more detailed breakdown of the results is provided in the following subsection.
On SpoofCeleb, the proposed \ModelName~substantially improves over both baselines while keeping all other configurations unchanged. These results indicate that the proposed cross-encoder-based \ModelName, combined with a simple three-class objective and LLR-based binary decision making, can effectively achieve comparable or even stronger SASV performance.

These contrasting results motivate a closer examination of what the proposed model learns and how its score distributions differ from those of existing baselines. In the following section, we analyze the class-conditional score patterns and the effect of operating-prior adjustment.

\subsection{Analysis of learned cues in the proposed model}

\begin{figure}[t]
    \centering
    \includegraphics[width=\linewidth]{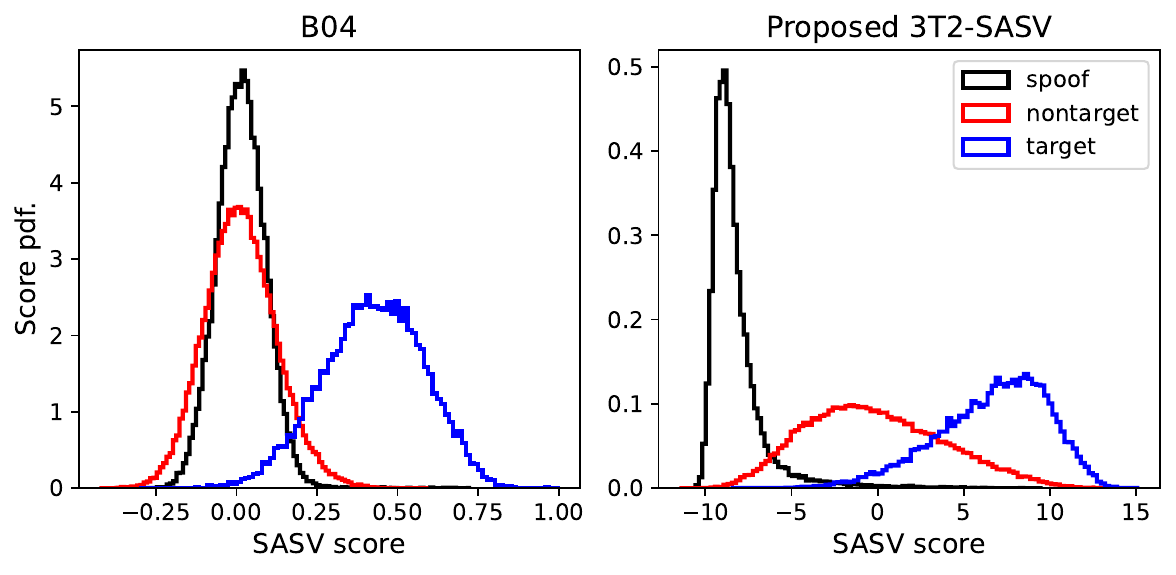}
     \vspace{-0.7cm}
    \caption{Comparison of score distributions between the ASV5-B04 baseline and the proposed~\ModelName~on SpoofCeleb.}
    \label{fig:score_distribution_spoofceleb}
    \vspace{-0.32cm}
\end{figure}

To further examine why the proposed model performs differently across datasets and understand the learned cues of the proposed model, we visualize the class-conditional score distributions.
 

As illustrated in Fig.~\ref{fig:score_distribution_spoofceleb}, the proposed model produces better-structured score distributions on SpoofCeleb: target trials concentrate in the high-score region, while nontarget and spoofed trials are pushed toward lower scores with clearer class-consistent structure. This aligns with the core motivation of our design. By modeling trial-level interactions and explicitly learning competing hypotheses of target, nontarget, and spoof, \ModelName~avoids compressing two rejection causes into a single similarity-like score, improving both separability and trial-level interpretability in an in-the-wild evaluation setting.

We observed a similar pattern in the ASVspoof~5 dataset that our proposed \ModelName~distinguishes spoof and nontarget trials better. But some residual overlap remains, caused by adversarial attacks (A18, A20, A23, A27, A30, and A31) and MaryTTS (A19). This indicates that, although the proposed model learns more discriminative class-conditional representations, the remaining overlap on a few difficult attacks may still limit improvements in the final min a-DCF. Improving the model's robustness to these attacks is left for future work.

\subsection{Adaptation to different operating priors}
The explicit three-class formulation in \ModelName~retains separate evidence for target, nontarget, and spoofed trials, allowing their contributions to the final LLR to be adjusted after training. We therefore investigate whether the proposed LLR can adapt to different deployment conditions by modifying only the class priors during inference, without retraining the model or recomputing the network outputs. This experiment is implemented on ASVspoof~5 with MFA-Conformer, it directly examines the claimed capability of post-hoc class-prior adjustment.

\begin{figure}[t]
    \centering
    \includegraphics[width=0.98\linewidth]{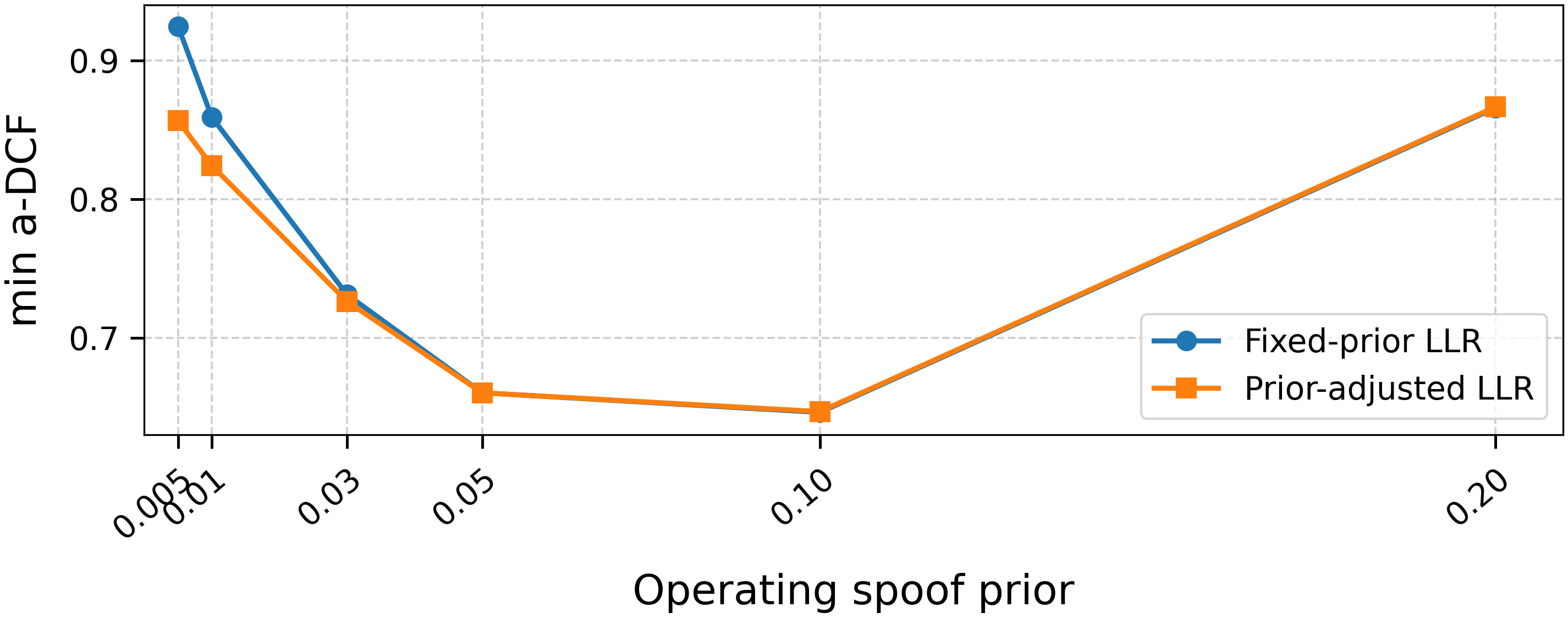}
    \vspace{-0.1cm}
    \caption{Min a-DCF across different operating spoof priors for the fixed-prior and prior-adjusted LLRs.}
    \label{fig:prior_adaptation_min_adcf}
    \vspace{-0.3cm}
\end{figure}

For each operating condition, the same priors are used in the a-DCF computation. We compare two inference strategies. The \emph{fixed-prior LLR} always uses the default inference priors $(\pi_{\mathrm{non}},\pi_{\mathrm{spf}})=(0.0095,0.05)$, whereas the \emph{prior-adjusted LLR} updates the priors used to combine the nontarget and spoof evidence according to the assumed operating condition.

Fig.~\ref{fig:prior_adaptation_min_adcf} reports min a-DCF values. At the default spoof prior of $0.05$, the two systems produce identical results, as expected. The benefit of prior adjustment is most evident under low-spoof-prior conditions. Specifically, when the operating spoof prior decreases to $0.005$, prior adjustment reduces min a-DCF from $0.9246$ to $0.8564$. As the spoof prior increases, the difference between the two strategies becomes negligible. Overall, the relative behavior of the two scoring strategies remains stable across the tested moderate-prior range, with greater sensitivity observed only under very low spoof priors. Since min a-DCF optimizes the threshold separately for each operating condition, the transferability of a single fixed threshold across different priors is not evaluated here.

These results show that the proposed three-class LLR permits the operating priors to be modified after training, particularly under a low spoof prior. 
The results therefore support the flexibility of the proposed \ModelName~for post-hoc prior adjustment.

\subsection{Extension to SSL-Based Embedding Extractors}
The previous experiments mainly evaluate \ModelName~with conventional SASV backbones, including MFA-Conformer and SKA-TDNN following~\cite{ASVspoof5}. To further examine whether the proposed three-class formulation generalizes beyond these conventional embedding extractors, we evaluate \ModelName~with an SSL-based embedding extractor.
Specifically, we use the default SSL-MHFA configuration in WeDefense~\cite{zhang2026wedefense}\footnote{\href{https://github.com/zlin0/wedefense/blob/main/egs/detection/asvspoof5/v15_ssl_mhfa/conf/MHFA_wav2vec2.yaml}{egs/detection/asvspoof5/v15\_ssl\_mhfa/conf/MHFA\_wav2vec2.yaml}} for embedding extractors, and integrate them into the proposed \ModelName~framework with following aggregation and classifier head.


To strengthen speaker-discriminative modeling, existing integrated SASV systems often adopt initialization from pretrained ASV models~\cite{kang22_interspeech, mun23_interspeech}, as in the baselines considered in this study. However, recent SSL models are pretrained on large-scale speech data and have been shown to encode rich speaker-related information. This motivates us to investigate whether the proposed \ModelName~framework can be instantiated from SSL-pretrained models and subsequently fine-tuned for SASV, without relying on ASV-pretrained initialization.



\begin{table}[t]
\centering
\caption{min a-DCF of wav2vec 2.0 and MFA-Conformer as embedding extractors for the proposed \ModelName.}
\label{tab:foundation models}
\begin{tabular}{lcr}
\toprule
Emb. Extractor & Dev.  & Eval.\\
\midrule
wav2vec 2.0 Base    & 0.2545    & 0.4821 \\
Wav2vec 2.0 Large   & 0.2293    & 0.4510 \\
MFA-Conformer       & 0.2126    & 0.5973 \\
\bottomrule
\end{tabular}
\vspace{-5mm}
\end{table}

As shown in Table~\ref{tab:foundation models}, MFA-Conformer achieves slightly better min a-DCF on the development set, whereas the SSL-based models perform better on the evaluation set. In particular, wav2vec 2.0 large yields the best evaluation result, reducing the min a-DCF from 0.5973 with MFA-Conformer to 0.4510. These results indicate that SSL foundation models generalize more effectively to the evaluation condition.

More importantly, these results demonstrate the robustness and flexibility of the proposed \ModelName~itself. Without requiring initialization from a pretrained ASV model, \ModelName~remains effective when paired with SSL-based embedding extractors, showing that its three-class formulation can successfully exploit speaker-discriminative information from modern self-supervised representations. This confirms that the proposed framework is not tied to a specific backbone type and can be readily extended to both conventional ASV systems and SSL foundation models.


\section{Conclusion}

We proposed \ModelName, an end-to-end integrated SASV framework that reformulates SASV as a three-class trial-level problem over target, nontarget, and spoof class. Unlike conventional SASV systems that collapse rejection evidence into a single score, the proposed method explicitly preserves class-level evidence and converts the three-class logits into an SASV LLR for binary accept/reject decision making. By introducing a cross-encoder architecture, \ModelName~also enables direct interaction between enrollment and test utterances, providing a more expressive trial representation.

Experimental results on ASVspoof~5 and SpoofCeleb show that the proposed framework achieves competitive performance on ASVspoof~5 and clear gains on SpoofCeleb over representative integrated baselines. Analysis of score distributions further indicates that the proposed three-class formulation better separates target, nontarget, and spoofed trials, improving interpretability of rejection decisions. In addition, the LLR-based inference allows post-hoc adjustment of class priors without retraining, which is particularly beneficial under low-spoof-prior operating conditions. We also verified that the proposed framework can be combined with SSL-based embedding extractors, demonstrating its flexibility across backbone types.
Future work will focus on improving robustness against more challenging spoofing attacks and cross-domain shifts. In particular, the remaining overlap among difficult spoof attacks suggests that stronger spoof-aware representations and improved calibration strategies may further enhance the generalization of the proposed framework.

\section{Acknowledgment}
The author would like to thank the organizer of WildSpoof.
This work was supported by the Office of the Director of National Intelligence (ODNI), Intelligence Advanced Research Projects Activity (IARPA), via the ARTS Program under contract D2023-2308110001. The work was partly supported by BUT IGA project no. FIT-S-23-8278.

\section{Generative AI Disclosure}
Generative AI tools were used to edit and polish the human-written draft, and design Figure 2. All AI-assisted text was reviewed by the authors before submission.

\balance
\bibliographystyle{IEEEtran}
\bibliography{main}
\newpage
\section{Appendix}
\subsection{Parameter selection}
\label{sec:ablation_studies}
\subsubsection{Different number of reference utterances}\label{sec:num_ref_utt}
In speaker verification systems, providing multiple reference utterances is commonly expected to offer richer speaker information and improve enrollment robustness. To examine whether this assumption holds in a cross-encoder-based SASV framework, we analyze the effect of the number of reference utterances on SASV performance. Specifically, we vary the number of reference utterances provided for each enrollment speaker while keeping the test utterance and all other experimental settings unchanged.
\begin{table}[th]
\centering
\caption{The Effect of Different Number of Ref Utterances}
\label{different num of ref utt}
\scalebox{.90}{
\begin{tabular}{lll}
\toprule
Num of Ref Utt & min a-DCF \\
\midrule
1  & 0.2126 \\
2  & 0.2337 \\
3  & 0.2586 \\
4  & 0.2366 \\
\bottomrule
\end{tabular}
}
\end{table}
Table~\ref{different num of ref utt} reports the performance under different numbers of reference utterances. Contrary to the common assumption that increasing enrollment data always improves performance, we observe that performance does not monotonically improve as the number of reference utterances increases. In particular, using a single reference utterance achieves the best performance, while using more than one reference utterance leads to a degraded performance overall.

This phenomenon suggests that simply increasing the number of reference utterances does not necessarily benefit SASV performance in a cross-encoder setting. One possible explanation is that incorporating multiple reference utterances introduces additional intra-speaker variability, which may complicate the joint modeling of reference and test utterances in a cross-encoder setting. As a result, the model may struggle to effectively attend to the most informative reference segments.\\
These results indicate that increasing the number of reference utterances does not necessarily benefit SASV performance in the proposed cross-encoder framework.

\subsubsection{Different utterance length}
Longer utterances are commonly assumed to provide richer acoustic information and improve speaker modeling. Here, we investigate the effect of utterance length on SASV performance. In this experiment, the duration of both reference and test utterances is varied, while other experimental settings remain unchanged.
\begin{table}[th]
\centering
\caption{The Effect of Different Utterance Length}
\label{different utt duration}
\scalebox{.90}{
\begin{tabular}{lll}
\toprule
Utt Length & min a-DCF \\
\midrule
2.0  & 0.2126 \\
3.0  & 0.2589 \\
5.0  & 0.2624 \\
\bottomrule
\end{tabular}
}
\end{table}
Table~\ref{different utt duration} presents the SASV performance under different utterance duration settings. The results show that the best performance is achieved with a short utterance duration of 2 seconds, while longer utterances consistently lead to degraded performance.This observation suggests that extending utterance duration may introduce additional acoustic variability or redundant information, which can complicate joint modeling in a cross-encoder-based SASV framework.

\subsubsection{Different number of attention heads}
Based on \ref{sec:aggregation_exp}, we further analyze the impact of the number of attention heads in the cross-attention aggregation strategy. In this experiment, all training and input configurations are fixed and only the number of attention heads is varied. Experiments are also conducted on the ASVspoof~5 development set using MFA-Conformer within the proposed \ModelName~ framework.
\begin{table}[th]
\centering
\caption{The effect of different number of attention heads on the ASVspoof5 development set.}
\label{different num of attention heads}
\scalebox{.90}{
\begin{tabular}{lll}
\toprule
Number of Attention Heads & min a-DCF \\
\midrule
1  & 0.2906 \\
2  & 0.2984 \\
4  & 0.2126 \\
8  & 0.2787 \\
\bottomrule
\end{tabular}
}
\end{table}

Table~\ref{different num of attention heads} reports the SASV performance with different numbers of attention heads. The results demonstrate that the number of attention heads has a significant effect on model performance. Specifically, using four attention heads achieves the best performance, with an a-DCF of 0.2126, while fewer and more heads lead to degraded performance.\\
This suggests that too few attention heads may limit modeling capacity, whereas too many heads may introduce unnecessary complexity, resulting in suboptimal performance. Therefore, four attention heads are used in the remaining experiments.

\subsection{Ablation studies}\label{sec:ablation_studies}

\subsubsection{Hard pairs vs. random sampling}
\label{sec:res:hard_vs_random}
As mentioned in \S\ref{sec:hard-pair}, the proposed \ModelName~ uses hard-pair sampling to construct trials during training when metadata are available. Random sampling remains a general strategy and is used when metadata are unavailable. Hard-pair sampling is expected to strengthen speaker-discriminative learning by exposing the model to more challenging enrollment-test pairs. This concern is consistent with recent ASV findings showing that sampling strategies can affect the extent to which models learn robust speaker-discriminative evidence~\cite{ren2025canllm}.


To evaluate the effect of hard-pair sampling,  we compare it with random sampling in the ASVspoof5 dataset.
Since ASVsoof5 only provides speaker gender in metadata, hard-pair nontarget trials are constructed by pairing different speakers of the same gender. This removes an obvious speaker-independent cue and yields more challenging nontarget pairs than unconstrained random sampling.
\begin{table}[t]
\centering
\caption{Comparison of sampling strategies on the development set of ASVspoof~5.}
\label{hard-pair strategy}
\begin{tabular}{lc}
\toprule
Strategy & min a-DCF \\
\midrule
Random sampling     & 0.2481 \\
Hard-pair   & 0.2126 \\
\bottomrule
\end{tabular}
\end{table}
Table~\ref{hard-pair strategy} compares the performance of random nontarget sampling and hard-pair nontarget sampling under the same training configuration. The results show that hard-pair sampling reduces the min a-DCF from 0.2481 to 0.2126, which suggests that constructing harder nontarget pairs helps strengthen speaker-discriminative learning.

\subsubsection{Different aggregation strategies}\label{sec:aggregation_exp}
SASV is inherently an enrollment-conditioned decision problem: the system must compare a test utterance against one or more enrollment utterances and determine whether the test utterance is bona fide and matches the claimed speaker. Accordingly, \ModelName~ operates on enrollment-test trials and explicitly models the interaction between enrollment and test utterances through the aggregation module. The resulting trial-level representation is expected to capture both enrollment-conditioned speaker-matching evidence and test-side spoofing cues, thereby aligning the learned representation with the target, nontarget, and spoof decision structure of SASV.

\begin{figure}[t]
    \centering
    \includegraphics[width=0.98\linewidth, trim={0 50 370 0}, clip]{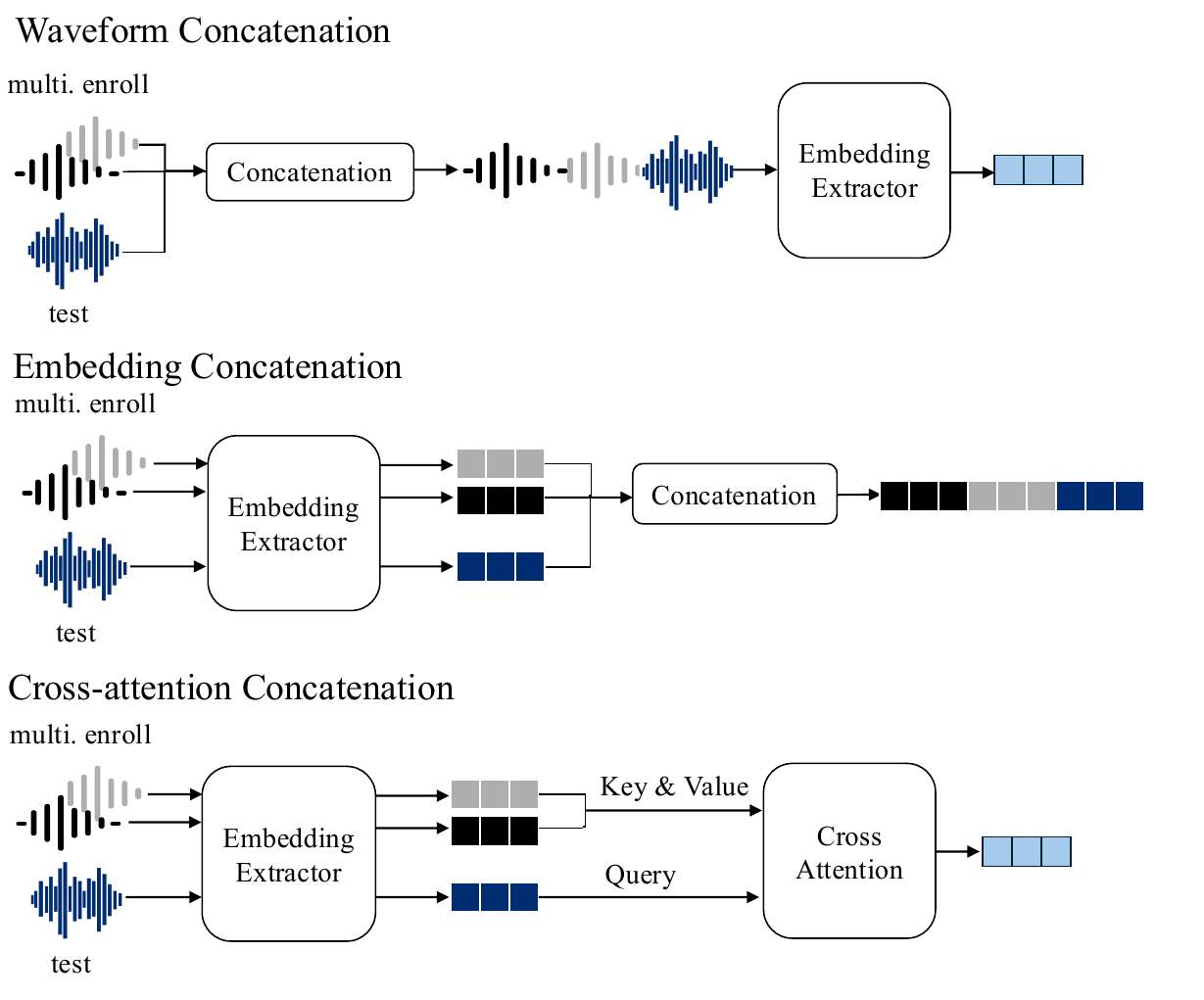}
    \vspace{-0.3cm}
    \caption{Comparison of three enrollment-test aggregation strategies in the proposed \ModelName~ framework.}
    \label{fig:aggregation_strategy}
    \vspace{-0.3cm}
\end{figure}

To evaluate the effectiveness of the cross-attention aggregation strategy in \ModelName~, we compared it with two simple aggregation strategies: waveform concatenation and embedding concatenation, as illustrated in Figure~\ref{fig:aggregation_strategy}. The proposed cross-attention strategy follows the formulation introduced in \S\ref{sec:fusion strategy}, where enrollment and test representations are interactively aggregated to construct a trial-level representation. In contrast, waveform concatenation combines the cropped enrollment and test waveforms at the input level and feeds the resulting $\mathbf{F}_t$ into the model to produce trial-level logits. Embedding concatenation first extracts utterance-level embeddings from the enrollment and test utterances independently, and then concatenates them before the final fully connected layer, leaving their interaction to be modeled only by the classifier head.

\begin{table}[t]
\centering
\caption{Effect of different enrollment-test aggregation strategies on the ASVspoof5 development set.}
\label{tab:different_fusion_strategy}
\begin{tabular}{lc}
\toprule
Aggregation Strategy & min a-DCF\\
\midrule
Embedding Concatenation   & 0.4071 \\
Waveform Concatenation  & 0.3297 \\
Cross-attention         & 0.2126 \\
\bottomrule
\end{tabular}
\end{table}
Experiments are conducted on the ASVspoof~5 development set using MFA-Conformer within the proposed \ModelName~ framework. The results are summarized in Table~\ref{tab:different_fusion_strategy}. As shown in the table, the proposed cross-attention aggregation consistently outperforms both waveform concatenation and embedding concatenation. This indicates that simply concatenating enrollment and test information is insufficient for modeling the trial-level structure of SASV. While concatenation-based methods combine the two inputs either at the waveform level or at the embedding level, their interaction is only implicitly learned by subsequent network layers. In contrast, cross-attention explicitly conditions the test representation on the enrollment embeddings, thereby capturing enrollment-test dependencies that are more directly aligned with the target, nontarget, and spoof decision structure.

\subsubsection{Effect of LLR calibration}
\label{sec:calibration_exp}

We further evaluate the calibrated LLR variant introduced in \S\ref{sec:llr}. Experiments are conducted on the ASVspoof~5 dataset using MFA-Conformer within the proposed 3T2-SASV framework. The calibration parameters are estimated on the development set, with $a=-0.149$, $b=3.918$, $c=0.783$, and $d=-0.102$. Although calibration is expected to improve score reliability when the development and evaluation score distributions are well matched, we observe that the calibrated variant increases the evaluation a-DCF by 0.012 compared with the uncalibrated LLR, as shown in Table~\ref{tab:llr_calibration}. This indicates that the calibration parameters learned on the development set do not transfer well to the evaluation condition, possibly due to a mismatch between development and evaluation score distributions. Therefore, we use the uncalibrated prior-adjusted LLR as the default scoring strategy in the remaining experiments.

\begin{table}[h]
\centering
\caption{Effect of LLR calibration on the ASVspoof~5 evaluation set.}
\label{tab:llr_calibration}
\scalebox{.90}{
\begin{tabular}{lc}
\toprule
Scoring strategy & min a-DCF \\
\midrule
Prior-adjusted LLR & 0.5741 \\
Calibrated LLR     & 0.5861 \\
\bottomrule
\end{tabular}
}
\end{table}

As shown in Figure~\ref{fig:score_distribution}, the proposed model still separates target, nontarget, and spoof trials to some extent on ASVspoof~5, but residual overlap remains, particularly for some spoof classes. To further investigate this confusing region, we break down the spoof scores by attack type for attacks that are difficult to distinguish from nontarget and target trials. As shown in Figure~\ref{fig:spoof_score_distribution}, the attack-wise score distributions suggest that the proposed model is more affected by several attack types, including adversarial attacks (A18, A20, A23, A27, A30 and 31) and MaryTTS (A19). One possible explanation is that these attacks may reduce stable spoofing artifacts or introduce cues that overlap with speaker-discriminative information. Consequently, these attacks are more likely to be assigned intermediate scores that overlap with the nontarget or target regions. Improving the model’s robustness to these attacks is left as future work.

\begin{figure}[t]
    \centering
    \includegraphics[width=\linewidth]{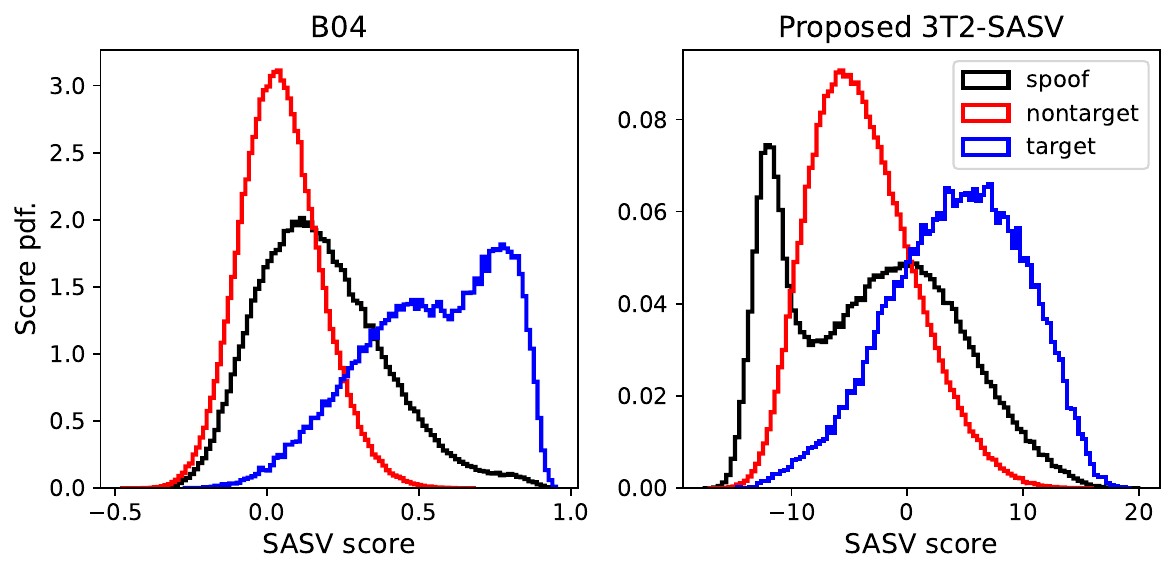}
    \vspace{-0.7cm}
    \caption{Class-conditional score distribution of the Integrated B04 method and the proposed method on ASVspoof~5.}
    \label{fig:score_distribution}
    \vspace{-0.5cm}
\end{figure}

\begin{figure}[t]
    \centering
    \includegraphics[width=0.98\linewidth]{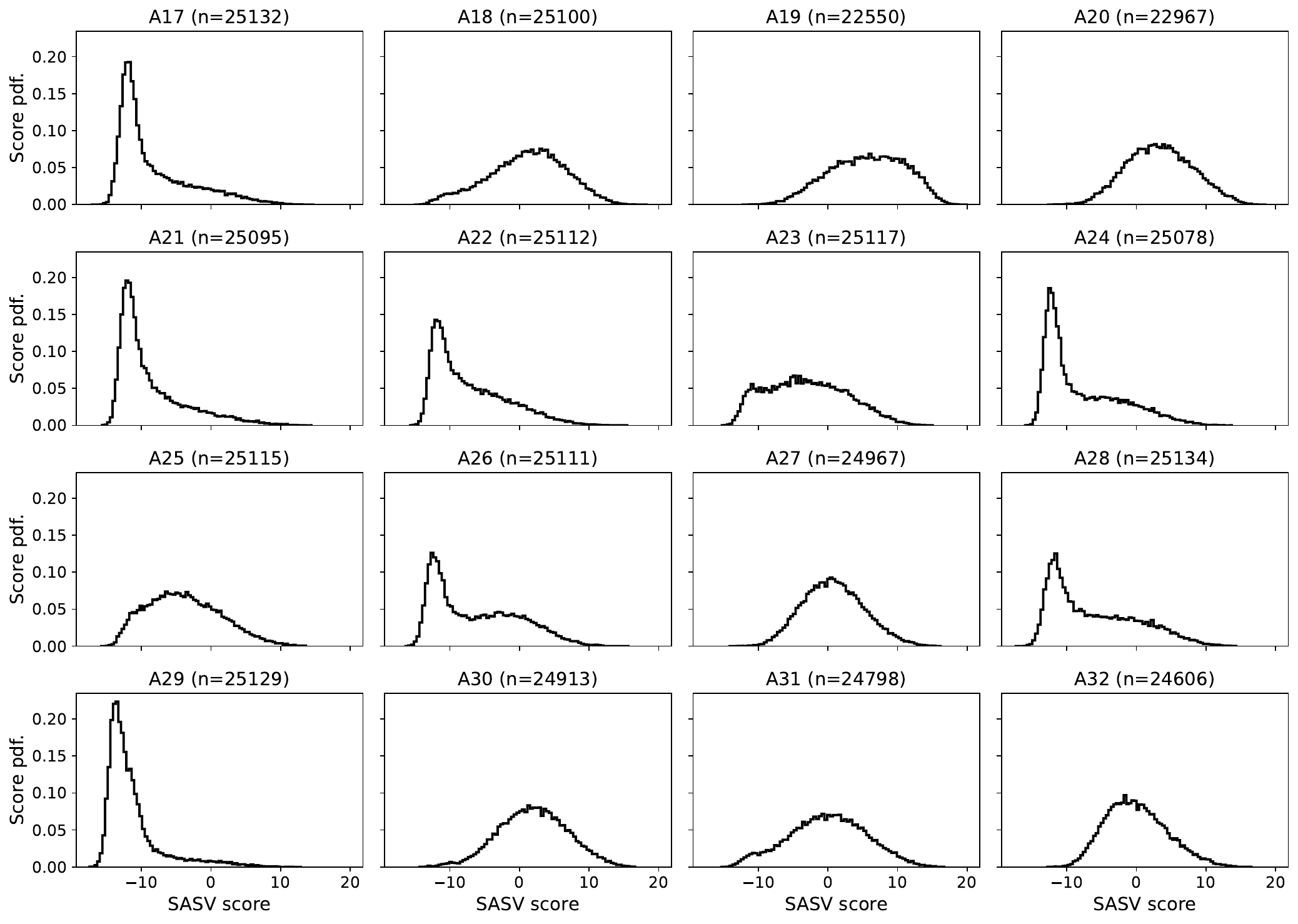}
    \vspace{-0.3cm}
    \caption{Attack-wise score distribution of the spoof class in the proposed method on ASVspoof~5.}
    \label{fig:spoof_score_distribution}
    \vspace{-0.3cm}
\end{figure}

\balance
\end{document}